\documentclass[uslettersize, 10pt, conference]{ieeeconf}      % Use this line for a4
\IEEEoverridecommandlockouts                              % This command is only
\usepackage{graphics} % for pdf, bitmapped graphics files
\usepackage{epsfig} % for postscript graphics files
\usepackage{mathptmx} % assumes new font selection scheme installed
\usepackage{times} % assumes new font selection scheme installed
\usepackage{amsmath} % assumes amsmath package installed
\usepackage{amssymb}  % assumes amsmath package installed
\usepackage{amscd}
\usepackage{cite}
\usepackage{euscript}
\usepackage{bbm}

\newtheorem{theorem}{Theorem}
\newtheorem{lem}[theorem]{Lemma}

\title{\LARGE \bf
A Remark on Formation Control with Triangulated Laman Graphs: \\ Genericity of Equivariant Morse Functions}

\author{Xudong Chen$^{1}$% <-this stops a space
\thanks{$^{1}$Xudong Chen is with Department of Electrical and Computer Engineering and Coordinated Science Laboratory, University of Illinois at Urbana-Champaign, email:
        {\tt\small xdchen@illinois.edu}}%
}

\begin{document}

\maketitle
\thispagestyle{empty}
\pagestyle{empty}

\begin{abstract} This paper, as a continuing work of \cite{chen2014reciprocal}, focus on establishing the fact that if we equip a reciprocal multi-agent (RMA) system with a triangulated Laman graph (TLG), then the associated potential function is generically an equivariant Morse function, i.e., there are only finitely many critical orbits each of which is nondegenerate. Though this assumption on the potential function of being an equivariant Morse function  has  been used, and in fact indispensable, in several occasions. But it is actually still an open question whether it is true for a given RMA system. Thus, in this paper we will provide a confirmative answer to the question for the class of RMA systems with TLGs. The main result, as well as the analysis of this paper, has many implications for other difficult problems.  
\end{abstract}

\section{Introduction}
In this paper, we will consider a special class of reciprocal multi-agent (RMA) systems equipped with a particular type of Laman graphs, as we call the triangulated Laman graphs. It is known that each RMA system is a gradient system with respect to an equivariant potential function, and in this paper we will develop among other things a basis property of this special class of RMA systems. We show that for each of such systems, the associated potential function is generically an equivariant Morse function, i.e, there are only finitely many critical orbits each of which is nondegenerate. 
\vspace{3pt}

We start by reviewing the mathematics model of a RMA system. Let $\mathbb{G} = (V,E)$ be an undirected graph of $N$ vertices, with $V= \{1,\cdots, N\}$ being the set of vertices and $E$ the set of edges. We say two vertices $i$ and $j$ are {\it adjacent} if $(i,j)$ is an edge of $\mathbb{G}$, and for convenience we let $V_i$ be the set of vertices which are adjacent to vertex $i$. Let $\{f_{ij}| (i,j)\in E\}$ be a family of continuous differentiable functions from $\mathbb{R}_+$, the set of positive real numbers, to $\mathbb{R}$. The equations of motion 
for a set of $N$ agents $\vec x_1,\cdots, \vec x_N$ are described by
\begin{equation}\label{MODEL}
\dot {\vec x}_i = \sum_{j\in V_i} f_{ij}(d_{ij}) \cdot (\vec x_j - \vec x_i), \hspace{10pt} \forall i = 1,\cdots,N 
\end{equation}
where $d_{ij}$ denotes the Euclidean distance between $\vec x_i$ and $\vec x_j$. As we see from the equation above,  the interaction between $\vec x_i$ and $\vec x_j$ is modeled by $f_{ij}$, and it depends only on the distance. Along this paper, we assume that $f_{ij} = f_{ji}$ for all $(i,j)\in E$, i.e, interactions among agents are reciprocal. 
\vspace{3pt}

It is known that each RMA system is a gradient system, with the potential function given by
\begin{equation}\label{PHI}
\Phi(\vec x_1,\cdots, \vec x_N):= \sum_{(i,j)\in E}\displaystyle\int^{d_{ij}}_1x f_{ij}(x) dx   
\end{equation}
It is  a summation of pairwise potentials between adjacent agents. Notice that the potential  function $\Phi$ depends only on relative distances among agents, so the value of $\Phi$ is invariant if we translate and/or rotate the whole configuration. In mathematics, it says that $\Phi$ is an equivariant function with respect to the group action of rigid motion. A precise definition will be given later in section III.
\vspace{3pt}

An open question left so far is to ask whether it is true, or at least generically true, that the potential function $\Phi$ is an equivariant Morse function, i.e, there are only finitely many critical orbits (orbits of equilibria) each of which is nondegenerate?    
This question is important not only because the mathematical statement itself, but also because it is an indispensable condition assumed in many developed theorems and their applications. For example, analyzing local exponential stabilities,  counting number of critical orbits \cite{UH2013E}, characterizing their regions of attractions, investigating system behaviors under perturbations \cite{AB2012CDC,sun2014CDC,USZB} and etc., all these questions rely on the assumption that $\Phi$ is an equivariant Morse function.

\vspace{3pt}

In this paper, we will prove that if the underlying graph $\mathbb{G}$ is a triangulated Laman graph (TLG), as we will define in the section III, then generically $\Phi$ will be an equivariant Morse function. We have showed in \cite{chen2014reciprocal} that a RMA system, equipped with a TLG, has several distinct properties. For example, for each equilibrium configuration of system \eqref{MODEL},  there is a geometric decomposition of the configuration into union of line sub-configurations each of which is also an equilibrium. Furthermore, we have developed a formula which relates the Morse-Bott index of a critical orbit, as an algebraic term, to this geometric decomposition.  In particular, the formula says that the Morse-Bott index of a critical orbit can be computed as the sum of the Morse-Bott  indices of critical orbits of these decomposed line sub-configurations. This formula has a potential impact on the design and control of RMA systems as it enables us to locate or place critical orbits with various Morse-Bott indices over the configuration space, we here refer readers to \cite{chen2014reciprocal} where we apply this formula to count and locate all the stable critical orbits within a special class of interaction laws.  
\vspace{3pt}

In this paper, we will follow the results achieved in paper \cite{chen2014reciprocal} to  develop several other relevant properties associated with this special class of RMA systems. After this introduction, we proceed as follows. In section II, we will introduce a class of interaction laws which will be considered in this paper, and  also we will  introduce the Whitney $C^1$-topology on it. In section III, we will review some key definitions about RMA systems, and then we will state the main theorem of this paper. This theorem claims that the potential function $\Phi$ is generically an equivariant Morse function. The rest of the paper is then devoted to the proof of this fact. In section IV, we will briefly review the canonical partition and the Morse-Bott index formula, as two key results developed in \cite{chen2014reciprocal}. In section V through VIII, we will mainly focus on critical line configurations. In particular, we will show in section VIII that generically, there are only finitely many critical orbits of line configurations each of which is nondegenerate. By combining this result with the Morse-Bott index formula, we will then be able to establish the genericity result.

\section{Class of monotone interaction laws} 
In this section, we will introduce a class of interaction laws, together with the $C^1$-Whitney topology on it.
\vspace{3pt}

Let $\mathbb{G}$ be an undirected graph of $N$ vertices. Let $P$ be the {\it configuration space} defined by
\begin{equation}
P:=\big \{(\vec x_1,\cdots,\vec x_N)\in \mathbb{R}^{2\times N}\big |\vec x_i\neq \vec x_j, \forall (i,j)\in E \big \} 
\end{equation} 
Configurations with collisions of adjacent agents are excluded. We will now define a class of interaction laws by which the solution of system \eqref{MODEL}, with any initial condition in $P$, exists for all time, and converges to the set of equilibria of system \eqref{MODEL}.
\vspace{3pt}

Let $\mathbb{R}_+$ be the set of positive real numbers, and let $C^1(\mathbb{R}_+,\mathbb{R})$  be the set of continuous differentiable functions from $\mathbb{R}_+$ to $\mathbb{R}$. For each function $f$ in $C^1(\mathbb{R}_+,\mathbb{R})$, we let 
\begin{equation}
\tilde f(d):=df(d)
\end{equation} 
We introduce $\tilde f$ because if $f$ is an interaction law between a pair of adjacent agents, then $\tilde f$ represents the actual magnitude of attraction/repulsion between them. We will use $f$ and $\tilde f$ in various occasions, and both of them are useful in this paper.  
\vspace{3pt}

Let $\mathcal{F}$ be a subset of  $C^1(\mathbb{R}_+,\mathbb{R})$ defined as follows. A function $f$ is in $\mathcal{F}$ if and only if
\vspace{2pt}
\begin{itemize}
\item[C1.]  $\tilde f'(d)>0$ for all $d>0$, and $\tilde f$ has a (unique) zero. 
\vspace{4pt}
\item[C2.]  $\displaystyle\lim_{d\to 0}\displaystyle\int^1_{d} \tilde f(x)dx=\infty$. 
\end{itemize}
\vspace{2pt}
We here note that the two functions $f$ and $\tilde f$ share the same zero, i.e, 
$f(d) = 0$ if and only if $\tilde f(d) = 0$. We impose these two conditions because the first condition implies that the interaction is a repulsion at a short distance, and an attraction at a long distance. The second condition prevents collisions of adjacent agents along the evolution, so then the solution of system \eqref{MODEL}, with any initial condition in $P$, exists for all time.    Moreover, we have showed in \cite{chenRMAS} that if each interaction law $f_{ij}$ satisfies conditions C1 and C2, then all critical orbits of system \eqref{MODEL} are contained in a compact subset of $P$. 
In this paper, we will assume that each interaction law $f_{ij}$ is in $\mathcal{F}$, thus we assume the  convergence of system \eqref{MODEL}. 
\vspace{3pt}

We will now equip this function space $\mathcal{F}$ with the Whitney $C^1$-topology. First we will describe the Whitney $C^1$-topology on $C^1(\mathbb{R}_+,\mathbb{R})$, and we describe it by defining a basis of open sets in $C^1(\mathbb{R}_+,\mathbb{R})$. Let $C^0(\mathbb{R}_+)$ be the set of continuous functions from $\mathbb{R}_+$ to $\mathbb{R}_+$. Then for each $f$ in $C^1(\mathbb{R}_+,\mathbb{R})$, and each $\delta$ in $C^0(\mathbb{R}_+)$, we define an open ball  $B_{\delta}(f)$ of $C^1(\mathbb{R}_+,\mathbb{R})$ by collecting any $g$ in $C^1(\mathbb{R}_+,\mathbb{R})$ such that 
\begin{equation}
|g(d) - f(d)| + |g'(d) - f'(d)| <\delta(d)
\end{equation}
for all $d>0$. By varying $f$ over $C^1(\mathbb{R}_+,\mathbb{R})$ and $\delta$ over $C^0(\mathbb{R}_+)$, we then get a basis of open sets in $C^1(\mathbb{R}_+,\mathbb{R})$. 
\vspace{3pt}

The subset $\mathcal{F}\subset C^1(\mathbb{R}_+,\mathbb{R})$ is then equipped with the subspace topology.  Notice that $\mathcal{F}$ is an open subset of $C^1(\mathbb{R}_+,\mathbb{R})$ with respect to the Whitney $C^1$-topology. So a subset $V$ of $\mathcal{F}$ is open in $\mathcal{F}$ if and only if it is open in $C^1(\mathbb{R}_+,\mathbb{R})$.  
\vspace{3pt}

Let $|E|$ be the cardinality of the edge set $E$, and  in the case where $\mathbb{G}$ is a Laman graph, we have $|E| = 2N -3$. The ensemble of interaction laws of system \eqref{MODEL} should be considered as an element in $\mathcal{F}^{|E|}$. In this paper, we equip $\mathcal{F}^{|E|}$ with the product topology, i.e, a set $U$ of $\mathcal{F}^{|E|}$ is open if and only if it is a union of $\Pi_{(i,j)\in E}V_{ij}$ with each $V_{ij}$ an open set in $\mathcal{F}$. Finally,  we say a property is {\it generic} if it holds on an open dense subset of $\mathcal{F}^{|E|}$.

\section{Definitions and the main theorem}
\begin{figure}[h]
\begin{center}
\includegraphics[scale=.35]{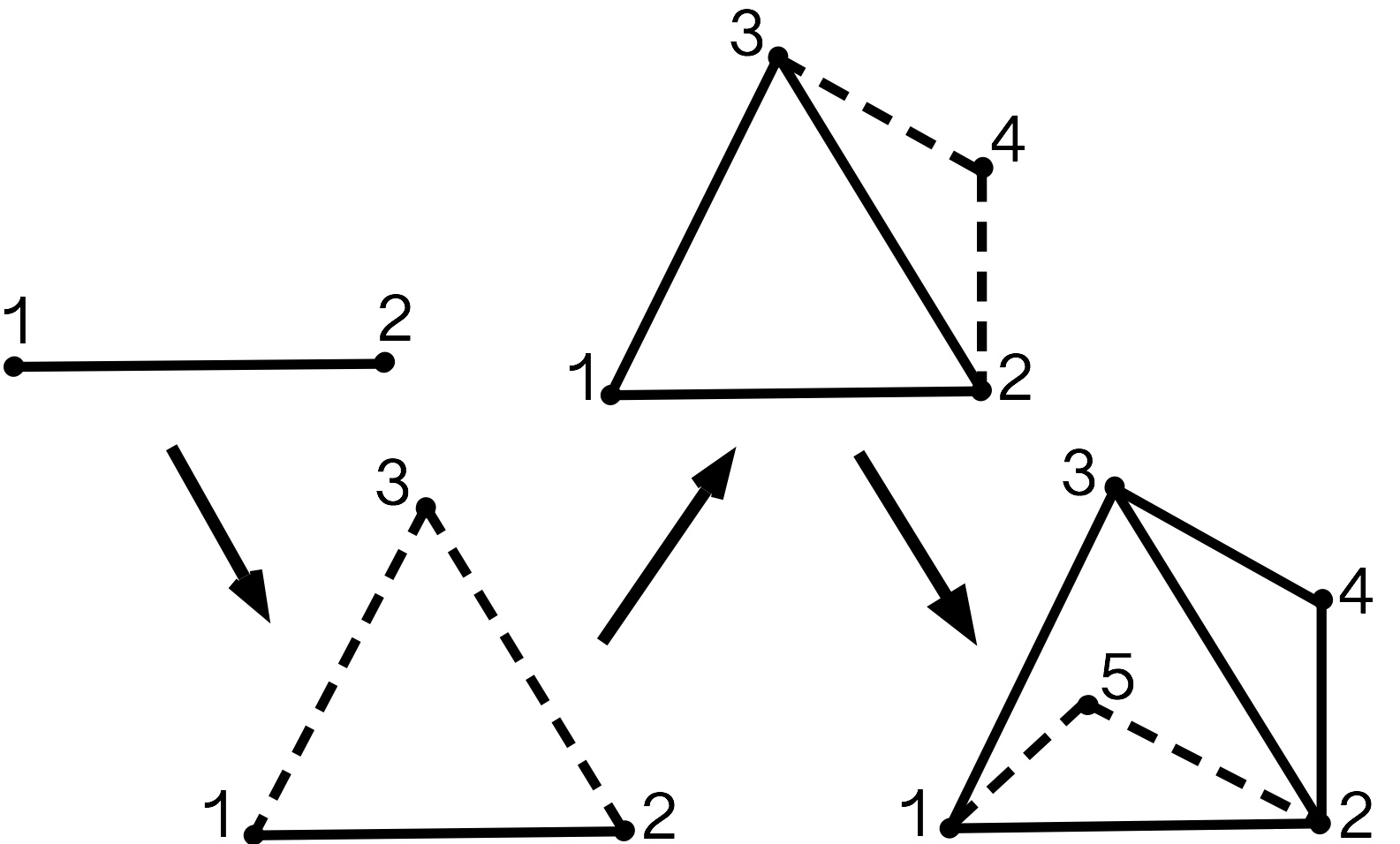}
\caption {An example of a TLG. Start with edge (1,2), we then subsequently join vertices 3, 4 and 5 to two existing adjacent vertices.}
\label{Hcon}
\end{center}
\end{figure}

In this section, we will review some definitions, as well as notations introduce in \cite{chen2014reciprocal}, and we state the main theorem of this paper. 
\vspace{5pt}
\\
{\it 1. Triangulated Laman graph}. Laman graphs are knowns as the minimally rigid graphs in $\mathbb{R}^2$ \cite{laman1970}, and each Laman graph can be constructed via a Henneberg construction. A triangulated Laman graph (TLG), as we will define now, can be constructed via a special Henneberg construction.  Start with an edge, we then join a new vertex, at each step, to two adjacent existing vertices via two new edges. An example of a TLG is illustrated in figure \ref{Hcon}.

\noindent
{\it 2. Group action of rigid motion}.  Let $SE(2)$ be the special Euclidean group for $\mathbb{R}^2$, each element $\gamma$ in $SE(2)$ can be represented by a pair $(\theta, \vec v)$ with $\theta$ in the special orthogonal group $SO(2)$ and $\vec v$ a vector in $\mathbb{R}^2$.  In this representation, the group multiplication of two elements $\gamma_1 = (\theta_1,\vec v_1)$ and $\gamma_2 = (\theta_2,\vec v_2)$ is given by 
$ 
\gamma_2\cdot \gamma_1  = (\theta_2 \theta_1, \theta_2 \vec v_1 + \vec v_2)
$. 
\vspace{3pt}

We now define a $SE(2)$-action on $P$ by sending  $\gamma = (\theta,\vec v)$ in $SE(2)$ and  $p=(\vec x_1,\cdots, \vec x_N)$ in $P$ to 
\begin{equation}
\gamma\cdot p:=(\theta \vec x_1+\vec v,\cdots,\theta \vec x_N+\vec v)
\end{equation} 
This group action is often referred as the group action of rigid motion because it preserves the shape of a configuration. In this paper, we let 
\begin{equation}
\mathcal{O}_p:= SE(2) \cdot p
\end{equation} be the orbit of $p$ with respect to the $SE(2)$-action. 
\vspace{5pt}
\\
{\it 3. Equivariant Morse function}. An important observation of the potential function $\Phi$ is that $\Phi$ depends only on relative distances between agents. Consequently we have
\begin{equation}
\Phi(p) = \Phi(\gamma\cdot p)
\end{equation} 
for any $p\in P$ and any $\gamma\in SE(2)$. In particular, if $p$ is an equilibrium of system \eqref{MODEL}, then so is $p'$ in $\mathcal{O}_p$.  In any of such case, we call $\mathcal{O}_p$ a {\it critical orbit}. 
\vspace{3pt}

Let $\mathcal{O}_p$ be a critical orbit, and let $H_p$ be the Hessian matrix of $\Phi$ at $p$, i.e, 
\begin{equation}
H_p:= \frac{\partial^2 \Phi(p)}{\partial p^2}
\end{equation}
Since interactions among agents are reciprocal, the Hessian matrix $H_p$ is thus symmetric, and hence all eigenvalues of $H_p$ are real. The null space of $H_p$ at least contains the tangent space of $\mathcal{O}_p$ at $p$. In other words, if we let $n_0(H_p)$ be the number of zero eigenvalues of $H_p$, then $n_0(H_p)$ is at least three. We say a critical orbit $\mathcal{O}_p$ is {\it nondegenerate} if  $n_0(H_p)$ is exactly three. (As the set of eigenvalues of $H_{p'}$ is invariant as $p'$ varies over $\mathcal{O}_p$, so this definition doesn't depend on the choice of $p$.)   
\vspace{3pt}

A potential function $\Phi$ is said to be an {\it equivariant Morse function} if there are only finitely many critical orbits of system \eqref{MODEL} each of which is nondegenerate.  We will now state the main theorem of this paper.  
\vspace{5pt}

\begin{theorem}\label{MAIN}
Let $\mathbb{G} = (V,E)$ be a TLG, and we assume that the ensemble of interaction laws of system \eqref{MODEL} is contained in $\mathcal{F}^{|E|}$. The potential function $\Phi$ defined by equation \eqref{PHI} is then generically an equivariant Morse function.   
\end{theorem}

\section{The canonical partition and \\
the Morse-Bott index formula}\label{MICF}
In this section, we will review some key results developed in \cite{chen2014reciprocal}.

\subsection{The canonical partition}

Let $\mathbb{G} = (V,E)$ be a TLG, and let $p$ be a configuration in $P$. We will now introduce the {\it canonical partition} of $E$ associated with $p$. Choose a Henneberg construction of $\mathbb{G}$, and we label the vertices with respect to the order of the construction. %Let $\mathbb{G}^k:=(V^k,E^k)$ be a subgraph of $\mathbb{G}$ by restricting $\mathbb{G}$ to vertices $V^k:=\{1,\cdots,k\}$. 
The partition is then defined inductively by following the construction.  
\vspace{3pt}
\\
\textit{Base case}. Start with the subgraph $\mathbb{G}'=(V',E')$ of $\mathbb{G}$ consisting of vertices $V'=\{1,2\}$. Since there is only one edge $(1,2)$ in $E'$, the partition of $E'$ is trivial. 
\vspace{3pt}
\\
\textit{Inductive step}. Suppose $\mathbb{G}'= (V',E')$ is a subgraph of $\mathbb{G}$ consisting of vertices $V'= \{1,\cdots,n-1\}$, and we have partitioned $E'$ into disjoint subsets as 
\begin{equation}
E' = E'_1\cup\cdots \cup E'_{m'}
\end{equation} 
Now suppose  vertex $n$ joins to vertices $i$ and $j$ via edges $(i,n)$ and $(j,n)$, and we describe the rule of updating the partition by taking into account $(i,n)$ and $(j,n)$.
\vspace{3pt}

Without loss of generality, we assume that the edge $(i,j)$ lies in $E'_1$,  then there are two cases:
\vspace{3pt}
\\
\textit{Case I}. If $\vec x_{i}$, $\vec x_j$ and $\vec x_n$ are aligned, then we update the partition by adding $(i,n)$ and $(j,n)$ into $E'_1$.
\vspace{3pt}
\\
\textit{Case II}. If $\vec x_i$, $\vec x_j$ and $\vec x_n$ are not aligned, then we update the partition  as follows
\begin{equation}
E'_1\cup\cdots\cup E'_{m'}\cup\{(i,n)\}\cup\{(j,n)\}
\end{equation}
By following the Henneberg construction, we then derive the canonical partition of $E$ associated with $p$.  
\vspace{3pt}

\begin{figure}[h]
\begin{center}
\includegraphics[scale=.35]{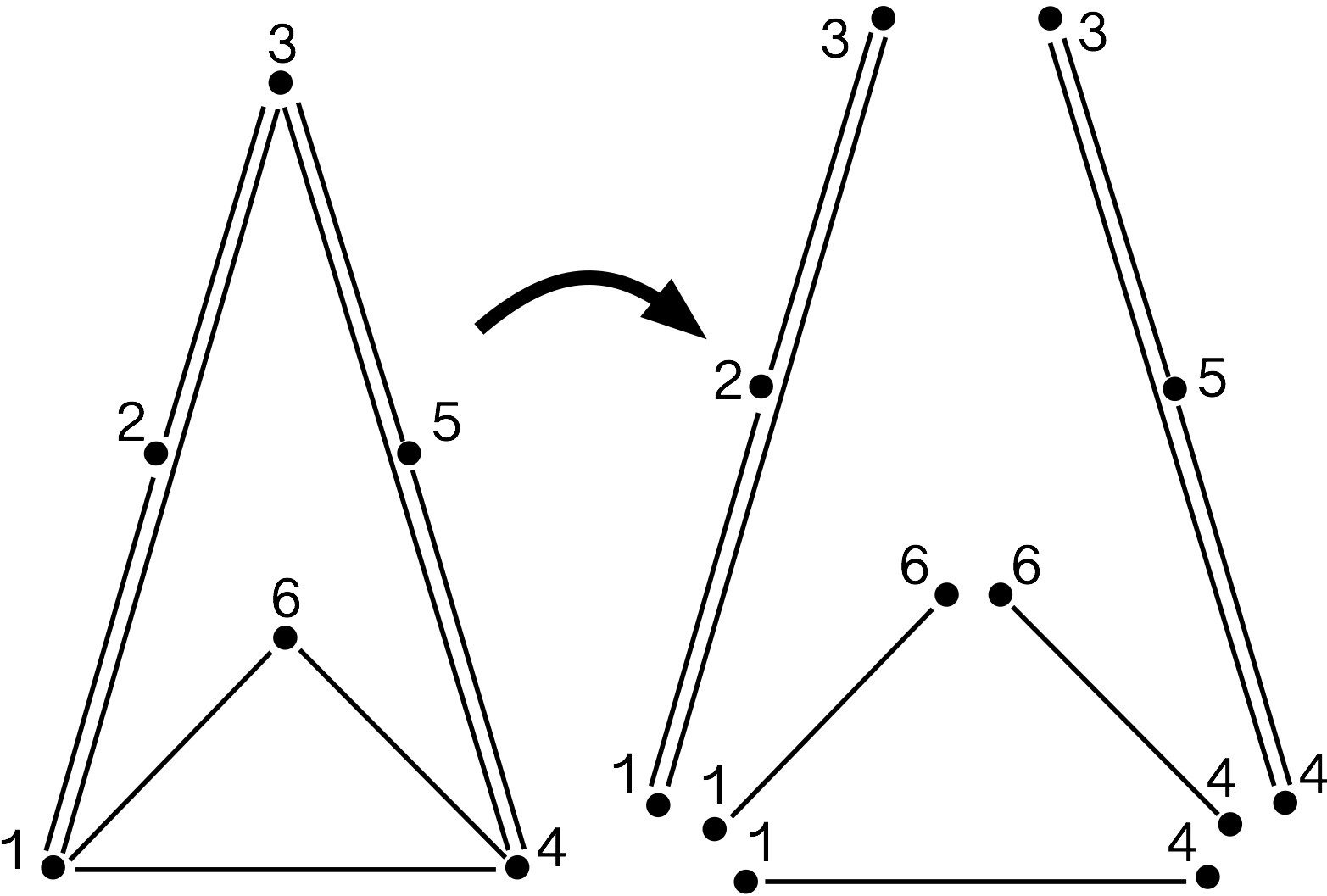}
\caption {An example of the canonical partition. We see from the left figure that the graph $\mathbb{G}$ is a TLG as we label the vertices with respect to a Henneberg construction, and $p$ is a planar configuration with $\vec x_1, \vec x_2,\vec x_3$ aligned, and $\vec x_3,\vec x_4,\vec x_5$ aligned. Then the canonical partition of $E$ associated with $p$ is given by the right figure.}
\label{DECO}
\end{center}
\end{figure}

We proved in \cite{chen2014reciprocal} that this partition does not depend on the choice of the Henneberg construction. An example  of the canonical partition is illustrated in figure \ref{DECO}.

\vspace{3pt}
We will now list some relevant properties of the canonical partition. Let $E=E_1\cup\cdots\cup E_m$ be the canonical partition associated with $p$. Let $\mathbb{G}_i=(V_i,E_i)$ be the subgraph of $\mathbb{G}$ by restricting $\mathbb{G}$ to $E_i$, and let $p_i$ be the sub-configuration of $p$ associated with $\mathbb{G}_i$, then  
\vspace{3pt}
\\
{a)}. Each $\mathbb{G}_i$ is a TLG.
\vspace{3pt}
\\
{b)}. Each $p_i$ is a line configuration. 
\vspace{3pt}
\\
{c)}. If there is another partition of $E$ satisfying  conditions a) and b), then it is a refinement of the canonical partition.  In other words, the canonical partition produces minimal number of sub-graphs regarding to the first two conditions. 
\vspace{3pt}
\\
{d)}. if $p$ is an equilibrium, then each $p_i$ is an equilibrium of the sub-system induced by $\mathbb{G}_i$. %In other words, if agents of $p_i$ form an isolated sub-system with  $\mathbb{G}_i$ the network topology and $\{f_{kl}|(k,l)\in E_i \}$ the set of interaction laws, then $p_i$ will be an equilibrium of the isolated sub-system. 
\vspace{3pt}
\\
More details, including the proofs of these statements, can be found in \cite{chen2014reciprocal}. 

\subsection{The Morse-Bott index formula}
In this part, we will follow the canonical partition to introduce a formula which can be used to compute the Morse-Bott index of a critical orbit. We start by introducing some necessary definitions.
\vspace{3pt}

Let $\mathcal{O}_p$ be a critical orbit, and let $H_p$ be the Hessian matrix of $\Phi$ at $p$. Let $n_+(H_p)$, $n_0(H_p)$, and $n_-(H_p)$ be the numbers of positive, zero, and negative eigenvalues of $H_p$ respectively. In this paper, we refer to the triplet $(n_+(M), n_0(M), n_-(M))$ as the {\it inertia} of  $H_p$. The {\it Morse-Bott index and co-index} of $\mathcal{O}_p$ are defined to be $n_-(H_p)$ and $n_+(H_p)$ respectively.  
\vspace{5pt}
\\
{\bf The Morse-Bott index formula}. Let $\mathbb{G}$ be a TLG, and let $\mathcal{O}_p$ be a critical orbit of system \eqref{MODEL}. Let $\{p_i\}^m_{i=1}$ and $\{\mathbb{G}_i\}^m_{i=1}$ be sub-configurations of $p$ and subgraphs of $\mathbb{G}$ respectively, associated with the canonical partition. Let $\Phi_i$ be the induced potential function associated with $\mathbb{G}_i$, i.e, 
\begin{equation}
\Phi_{i}(p'_i) := \sum_{(j,k)\in E_i} \displaystyle \int^{d_{jk}}_1 \tilde f_{jk}(x) dx 
\end{equation} 
and let $H_{p_i}$ be the Hessian matrix of $\Phi_i$ at $p_i$.  Then we have
\begin{equation}\label{INDEXF}
\left\{
\begin{array}{l}
n_-(H_p)=\sum^m_{i=1}n_-(H_{p_i}) \vspace{3pt}\\
n_+(H_p)=\sum^m_{i=1}n_+(H_{p_i})
\end{array}
\right.
\end{equation}
This set of equations will be referred as the Morse-Bott index formula. This formula has some relevant implications as we state below
\begin{itemize}
\item[1).] The critical orbit $\mathcal{O}_p$ is nondegenerate if and only if each $\mathcal{O}_{p_i}$ is nondegenerate. 
\vspace{3pt}

\item[2).] Suppose $\mathcal{O}_p$ is nondegenerate, then $\mathcal{O}_p$ is (exponentially) stable, i.e, $n_+(H_p) = 0$ if and only if each $\mathcal{O}_{p_i}$ is (exponentially) stable.  
\end{itemize}
\vspace{3pt}
We refer readers to  \cite{chen2014reciprocal} for a complete proof of the formula, as well as the proofs of the implications above.

\section{The Hessian matrix at a line configuration}\label{HMLC}
As we have seen in the last section that the Morse-Bott index of $\mathcal{O}_p$ can be computed as the sum of Morse-Bott indices of $\mathcal{O}_{p_1},\cdots,\mathcal{O}_{p_m}$ with $p_1,\cdots,p_m$ the line sub-configurations of $p$ associated with the canonical partition. So in this section, we will focus on the case where $p$ itself is a critical line configuration, and we will compute the Hessian of $\Phi$ at $p$. 
\vspace{3pt}

In the rest of this paper, we assume that the $a$-axis and the $b$-axis are the two axes of $\mathbb{R}^2$, and we let $a_i$ and $b_i$ be the two coordinates of $\vec x_i$. Let %$\vec a$ and $\vec b$ be two vectors in $\mathbb{R}^N$ collecting the $a$-coordinates and $b$-coordinates of agents respectively, i.e, 
\begin{equation}
\left\{
\begin{array}{ll}
\vec a := (a_1,\cdots,a_N)\vspace{3pt}\\ 
\vec b := (b_1,\cdots,b_N)
\end{array}
\right.
\end{equation}
We then re-arrange entries of a configuration $p$ so that 
\begin{equation}
p = (\vec a,\vec b)
\end{equation}
The Hessian matrix $H_p$ will then be computed with respect to this order.  
\vspace{3pt}

Since the only information we need in this paper is the set of eigenvalues of  $H_p$, and since the set of eigenvalues of $H_{p'}$ is invariant as $p'$ varies over the orbit $\mathcal{O}_p$,  we may rotate and/or translate $p$ if necessary so that the entire configuration $p$ is on the $a$-axis.  
\vspace{3pt}

To compute $H_p$, we will now introduce two $N$-by-$N$ matrices $F_p$ and $d\tilde F_p$. Both of these two matrices are symmetric, of zero-row/column-sum. So we define these two matrices by specifying their off-diagonal entries. Let $F_{p,ij}$ be the $ij$-th entry of $F_p$, and $d\tilde F_{p,ij}$ the $ij$-th entry of $d\tilde F_p$,  we then define
\begin{equation}\label{GMAT}
F_{p,ij}:=
\left\{
\begin{array}{ll}
f_{ij}(d_{ij}) & \text{if }(i,j)\in E\vspace{3pt}\\
0 & \text{otherwise}
\end{array}
\right.
\end{equation}
and       
\begin{equation}
d\tilde F_{p,ij}:=
\left\{
\begin{array}{ll}
\tilde f'_{ij}(d_{ij}) & \text{if }(i,j)\in E\vspace{3pt}\\
0 & \text{otherwise}
\end{array}
\right.
\end{equation}
The Hessian of $\Phi$ at $p$ is then given by
\begin{equation}\label{HESS}
H_p =
\begin{pmatrix}
d\tilde F_p & 0\\
0 & F_p
\end{pmatrix}
\end{equation}
We will now discuss about the null space of $H_p$. The null space of $H_p$ at least contains $T_p\mathcal{O}_p$, i.e, the tangent space  of $\mathcal{O}_p$ at $p$, as we will compute now.  
\vspace{3pt}

Let $\vec e$ be a vector in $\mathbb{R}^N$ of all ones, and we define two vectors in $\mathbb{R}^{2\times N}$ by
\begin{equation}
\left\{
\begin{array}{ll}
\vec t_a: = (\vec e, 0)\vspace{3pt}\\
\vec t_b: = (0,\vec e)
\end{array}
\right.
\end{equation} 
These two vectors represent the infinitesimal motions of translation along $a$-axis and $b$-axis respectively. We now define another vector in $\mathbb{R}^{2\times N}$ by
\begin{equation}
\vec r_p: = (0, \vec a)
\end{equation} 
This vector then represents the infinitesimal motion of clockwise rotation of $p$ around the origin. 
\vspace{3pt} 

These three vectors $\vec t_a$, $\vec t_b$ and $\vec r_p$ form a basis of the tangent space $T_p\mathcal{O}_p$. By direct computation, we verify that all the three vectors $\vec t_a$, $\vec t_b$ and $\vec r_p$ are in the null space of $H_p$. On the other hand, if the critical orbit $\mathcal{O}_p$ is nondegenerate, then the null space of $H_p$ coincides with $T_p\mathcal{O}_p$. In particular, we note that if $\mathcal{O}_p$ is nondegenerate, then the null space of $d\tilde F_p$ should only be spanned by $\vec e$, and the null space of $F_p$ should only be spanned by $\vec e$ and $\vec a$.   

%%%%%%%%%%%%%%%%%%%%%%%%%%%%%%%%%%%%%%%%%%%%%%%%%%%%%%%%%
%%%%%%%%%%%%%%%%%%%%%%%%%%%%%%%%%%%%%%%%%%%%%%%%%%%%%%%%%
%%%%%%%%%%%%%%%%%%%%%%%%%%%%%%%%%%%%%%%%%%%%%%%%%%%%%%%%%

\section{The reduced system}\label{VIL}
In this section, we will introduce the notion of reduced system which will be a useful tool for analyzing and computing the inertia of the Hessian matrix at a critical line configuration.   
\vspace{3pt}

We start by introducing the notion of virtual interaction. Consider an auxiliary system of three agents $\vec x_1$, $\vec x_2$ and $\vec x_3$. Let $f_{12}$ and $f_{13}$ be interaction laws between $\vec x_1$ and $\vec x_2$, and between $\vec x_1$ and $\vec x_3$ respectively.  We assume that  $\vec x_1$, $\vec x_2$ and $\vec x_3$ are aligned on the $a$-axis, and we will now define the virtual interaction between $\vec x_2$ and $\vec x_3$ induced by $f_{12}$ and $f_{13}$. There are three cases, depending on which of the three agents lies in between the other twos. 
\vspace{5pt}
\\
{\it Case 1}. Agent $\vec x_1$ lies in between $\vec x_2$ and $\vec x_3$, and without loss of generality, we assume that the three coordinates $a_1$, $a_2$ and $a_3$ satisfy the condition that $a_2<a_1<a_3$. As both $\tilde f_{13}$ and $\tilde f_{13}$  are monotonically increasing and have zeros, so if we fix agent $\vec x_2$ and $\vec x_3$ but move $\vec x_1$ along the $a$-axis with $a_1\in (a_2,a_3)$, then there is a unique position for agent $\vec x_1$ at which it is balanced, i.e, 
\begin{equation}
f_{12}(d_{12})\cdot (a_1 - a_2) = f_{13}(d_{13})\cdot (a_3 - a_1)
\end{equation}
We then define a map $g_{23}: \mathbb{R}_+\rightarrow \mathbb{R}$ by requiring 
\begin{equation}
\begin{array}{lll}
g_{23}(d_{23})\cdot (a_3 - a_2) & =&  f_{12}(d_{12})\cdot (a_1 - a_2) \vspace{3pt}\\
& = &  f_{13}(d_{13})\cdot (a_3 - a_1)
\end{array}  
\end{equation}
and we call $g_{23}$ the {\it virtual interaction} between agents $\vec x_2$ and $\vec x_3$ induced by $ f_{12}$ and $f_{13}$. We here note that in this case, the map $g_{23}$ is a function in $\mathcal{F}$. To see this, we first notice that $g_{23}$ is continuous differentiable, with its derivative given by
\begin{equation}
\tilde g'_{23}(d_{23}) = \frac{\tilde f'_{12}(d_{12})\tilde f'_{13}(d_{13})}{\tilde f'_{12}(d_{12}) + \tilde f'_{13}(d_{13})}>0
\end{equation}
We then notice that  
\begin{equation}
\tilde g_{23}(d_{23}) = \tilde f_{12}(d_{12}) < \tilde f_{12}(d_{23}) 
\end{equation}
and hence 
\begin{equation}
\lim_{d\to 0} \displaystyle\int^1_d \tilde g_{23}(x) dx < \lim_{d\to 0}\displaystyle \int^1_d \tilde f_{12}(x) dx = -\infty 
\end{equation}
So the map  $ g_{23}$ satisfies both conditions C1 and C2, and hence $\tilde g_{23}$ is contained in $\mathcal{F}$. 
\vspace{5pt}
\\
{\it Case 2}. Agent $\vec x_2$ lies in between $\vec x_1$ and $\vec x_3$, and we assume  $a_1<a_2<a_3$. Similarly, if we fix $\vec x_2$ and $\vec x_3$ but move $\vec x_1$ along the $a$-axis with $a_1\in (-\infty,a_2)$, then there is a unique position for agent $\vec x_1$ at which it is balanced. Again, we have  
\begin{equation}
f_{12}(d_{12})\cdot (a_1 - a_2) = f_{13}(d_{13})\cdot (a_3 - a_1)
\end{equation}
and we define the virtual interaction $g_{23}: \mathbb{R}_+\rightarrow \mathbb{R}$ by requiring
\begin{equation}
\begin{array}{lll}
g_{23}(d_{23})\cdot (a_3 - a_2) & =&  f_{12}(d_{12})\cdot (a_1 - a_2) \vspace{3pt}\\
& = &  f_{13}(d_{13})\cdot (a_3 - a_1)
\end{array}  
\end{equation}
Similarly, the derivative of $\tilde g_{23}$ is given by  
\begin{equation}
\tilde g'_{23}( d_{23}) = \frac{\tilde f'_{12}(d_{12})\tilde f'_{13}(d_{13})}{\tilde f'_{12}(d_{12}) + \tilde f'_{13}(d_{13})}>0 
\end{equation}
but in this case $\lim_{d\to 0}\tilde g_{23}(d)$ is finite. To see this, we notice that both $\tilde f_{12}$ and $\tilde f_{13}$ are strictly increasing, so there is a unique $d_0>0$ such that 
\begin{equation}
\tilde f_{12}(d_0) + \tilde f_{13}(d_0) = 0 
\end{equation}
but then, we have 
\begin{equation}
\lim_{d\to 0}\tilde g_{23}(d) = \tilde f_{13}(d_0) > -\infty 
\end{equation}
We note that although $g_{23}$ is not in $\mathcal{F}$, yet $f+g_{23}$ will be in $\mathcal{F}$ for any $f$  in $\mathcal{F}$.
\vspace{5pt}
\\
{\it Case 3}. Agent $\vec x_3$ lies in between $\vec x_1$ and $\vec x_2$. We then follow the same procedure as we did in case 2 to construct the virtual interaction $g_{23}$. Again, in this case $g_{23}$ is not contained in $\mathcal{F}$, but still we have  $\tilde g'_{23}(d)>0$ for all $d>0$. 
\vspace{5pt}

Equipped with the notion of virtual interaction, we will then be able to introduce the reduced system. Let $\mathbb{G}=(V,E)$ be a TLG of $N$ vertices. Choose a Henneberg construction of $\mathbb{G}$, and we assume, along the rest of this paper, that vertex $1$ is the last vertex joining to the graph,   via edges $(1,2)$ and $(1,3)$ to vertices $2$ and $3$. Let $g_{23}$ be the virtual interaction between $\vec x_2$ and $\vec x_3$ induced by $ f_{12}$ and $ f_{13}$ defined in any of the three cases. We then let 
\begin{equation}\label{fstar}
 f^*_{23}(d):=f_{23}(d) +  g_{23}(d)
\end{equation}
and it is clear that $f^*_{23}$ is contained in $\mathcal{F}$. We then define a new system of $(N-1)$-agents by ruling out agent $\vec x_1$, and meanwhile replacing $ f_{23}$ with $f^*_{23}$. This newly defined  system will be referred as a {\it reduced system} in the rest of this paper. As there are three different ways to define $g_{23}$, and hence $f^*_{23}$, so there will be three types of reduced systems.  
\vspace{3pt}

As later in this paper, we will consider variations of interaction laws in both the original system and one of its reduced systems. So we find it helpful to introduce a map describing the relation between the two ensembles of interaction laws. For convenience, we let 
\begin{equation}
\left\{
\begin{array}{l}
\Omega: = \mathcal{F}^{2N-3} \vspace{3pt}\\
\Omega^* := \mathcal{F}^{2N - 5}
\end{array}
\right.
\end{equation}
 be the collections of  ensembles of interaction laws of the original system and a reduced system, respectively.  We then consider maps 
 \begin{equation}
 \rho_i:\Omega \to \Omega^*
 \end{equation} 
 with $i=1,2,3$, each of which is defined by 
\begin{equation}
\rho_i(f_{12},f_{13}, f_{23}, \cdots ) := (f^*_{23},\cdots)
\end{equation}
The sub-index $i$ of $\rho_i$ indicates in which case the map $f^*_{23}$ is defined. We here note  that each map $\rho_i$ is open, surjective and continuous, and we refer readers to the appendix for a complete proof of this fact. 
\vspace{5pt}

The notion of reduced system will be a useful tool for computing the inertia of the Hessian matrix of a critical orbit, and for analyzing generic properties associated with the set of critical orbits. In particular, it enables us to apply the technique of induction as we will see later in this paper.

\section{The inertia of the Hessian matrix\\ associated with the reduced system}
In this section, we will describe a useful property of reduced system which is related to the inertia of the Hessian matrix at a critical line configuration. 
\vspace{3pt}

We start by introducing some useful definitions and notations. Let $p$ be a fixed line configuration,  and we say a reduced system is associated with $p$ if the virtual interaction $g_{23}$ is defined  with respect to the arrangement of positions of the three agents $\vec x_1$, $\vec x_2$ and $\vec x_3$ in $p$. Let $M$ be a  symmetric matrix, and we will let 
\begin{equation}
\vec n(M):=(n_+(M), n_0(M), n_-(M))
\end{equation} 
be the inertia of $M$. For convenience, we also define a vector-valued sign function by 
\begin{equation}
sgn(x):=\left\{
\begin{array}{ll}
(1,0,0) & \text{ if } x > 0\\
(0,1,0) & \text{ if } x = 0\\
(0,0,1) & \text{ if } x < 0
\end{array}\right.
\end{equation} 
We will now state the property of the reduced system associated with $p$ 
\vspace{5pt}

\begin{theorem}\label{ASThm1} 
Suppose $p$ is a critical line configuration, then the sub-configuration $p^*$, formed by agents $\vec x_2,\cdots, \vec x_{n}$, is a critical line configuration of the reduced system associated with $p$. Let $\Phi^*$ be the associated potential function of the reduced system, and let $H_{p^*}$ be the Hessian matrix of $\Phi^*$ at $p^*$, then we have
\begin{equation}
\begin{array}{lll}
\vec n(H_p) - \vec n(H_{p^*}) & = & sgn(-f_{12}(d_{12}) - f_{13}(d_{13}))  \vspace{3pt}\\
& & +sgn(-\tilde f'_{12}(d_{12})- \tilde f'_{13}(d_{13}))
\end{array}
\end{equation} 
The expression above we be referred as the inertia formula in the rest of this paper. 
\end{theorem}
\vspace{5pt}

It is clear by construction of the virtual interaction that the sub-configuration $p^*$ is an equilibrium of the reduced system. So in the rest of this section, we will focus on the proof of inertia formula.  We will assume that the line configuration $p $ is on the $a$-axis. By previous computation, we have
\begin{equation}
H_p = 
\begin{pmatrix}
d\tilde F_p & 0\\
0 & F_p
\end{pmatrix}
\end{equation}
Let $F_{p^*}$ and $d\tilde F_{p^*}$ be two matrices defined in the same way as $F_p$ and $d\tilde F_p$, yet with respect to the reduced system associated with $p$. Then we have
\begin{equation}
H_{p^*} = 
\begin{pmatrix}
d\tilde F_{p^*}& 0\\
0 & F_{p^*} 
\end{pmatrix}
\end{equation}
It then suggests that we relate  $\vec n(F_p)$ and $\vec n(d\tilde F_p)$ to  $\vec n(F_{p^*})$ and  $\vec n(d\tilde F_{p^*})$, respectively.  The results are then summarized in the statements of lemma \ref{ASlem1} and lemma \ref{ASlem2}. 
\vspace{5pt}  

\begin{lem}\label{ASlem1}
The inertia of $F_p$ and $F_{p^*}$ are related by 
$
\vec n(F_p) - \vec n(F_{p^*}) = sgn(-f_{12}(d_{12}) - f_{13}(d_{13}))
$. 
\end{lem}  
\vspace{5pt}

\begin{proof} Let $\vec v^*_1,\cdots,\vec v^*_{N-1} $ be a set of orthonormal eigenvectors of $F_{p^*}$ with respect to eigenvalues $\lambda_1,\cdots,\lambda_{N-1}$. We will now use each $\vec v^*_i$ to construct a vector $\vec v_i$ in $\mathbb{R}^N$. Let $\vec v^*_{ij}$ be the $j$-th entry of vector $\vec v^*_i$, and as usual we let $a_k$ be the coordinate of agent $\vec x_k$ on the $a$-axis. We then define $\vec v_i:=  (\alpha,\vec v^*_i)$  by adding a scalar $\alpha$ in front of $\vec v^*_i$, and the scalar $\alpha$ is given by 
\begin{equation}\label{vecu2}
\alpha:=\frac{(a_3-a_1)v^*_{i1}+(a_1-a_2)v^*_{i2}}{a_3-a_2}
\end{equation}
Notice that in any of the three cases, the virtual interaction $g_{23}$ satisfies the condition
\begin{equation}\label{VIEq1}
\begin{array}{lll}
g_{23}(d_{23})\cdot (a_3 - a_2) & =&  f_{12}(d_{12})\cdot (a_1 - a_2) \vspace{3pt}\\
& = &  f_{13}(d_{13})\cdot (a_3 - a_1)
\end{array}  
\end{equation}
This, in particular, implies that
\begin{equation}\label{ASEq1}
F_p\cdot \vec v_i = \lambda_i \cdot (0,\vec v^*_i)
\end{equation}
for all $i=1,\cdots, N-1$.
\vspace{3pt}

Let $\vec e_1$ be a unit vector with one on its first entry, and zeros elsewhere.  Let $Q$ be a $N$-by-$N$ matrix, with $\vec e_1,\vec v_1,\cdots,\vec v_{N-1}$ its column vectors, then $Q$ is a full-rank matrix. We then consider a congruence transformation of $F_p$ by
\begin{equation}
\Lambda:= Q^TF_p Q
\end{equation} 
By expression \eqref{ASEq1}, we know that $\Lambda$ is a diagonal matrix  given by 
\begin{equation}
\Lambda= diag(-f_{12}(d_{12}) -f_{13}(d_{13}), \lambda_1,\cdots,\lambda_{N-1} )
\end{equation}
Then by applying the Sylvester's law of inertia \cite{sylvester1852xix}, we  know that the inertia of $F_p$ coincides with the inertia of $\Lambda$.  This then proves the lemma. 
\end{proof}
\vspace{5pt}

\begin{lem}\label{ASlem2}
The inertia of $d\tilde F_p$ and $d\tilde F_{p^*}$ are related by 
$
\vec n(d\tilde F_p) - \vec n(d\tilde F_{p^*}) = sgn(-\tilde f'_{12}(d_{12}) - \tilde f'_{13}(d_{13}))
$. 
\end{lem}  
\vspace{5pt}

\begin{proof}
The proof here will be very similar to the proof of lemma \ref{ASlem1}. Let $\vec u^*_1,\cdots,\vec u^*_{N-1}$ be a set of orthonormal eigenvectors of $d\tilde F_{p^*}$ with respect to eigenvalues $\tilde \lambda_1,\cdots,\tilde \lambda_{N-1}$.  Similarly, we  let $\vec u_i := (\beta, \vec u^*_i) $ with 
\begin{equation}
\beta:= \frac{\tilde f'_{12}(d_{12})u^*_{i1} + \tilde f'_{13}(d_{13})u^*_{i2} }{\tilde f'_{12}(d_{12}) + \tilde f'_{13}(d_{13})}
\end{equation}
where $u^*_{i1}$ and $u^*_{i2}$ are the first and second entries of $\vec u^*_i$ respectively.   Notice that 
\begin{equation}
\tilde g'_{23} (d_{23}) =  \frac{\tilde f'_{12}(d_{12})\tilde f'_{13}(d_{13})}{\tilde f'_{12}(d_{12})+ \tilde f'_{13}(d_{13})} 
\end{equation}
So then, we have 
\begin{equation}
d\tilde F_p \cdot \vec u_i=\tilde \lambda_i\cdot (0,\vec u^*_i) 
\end{equation}
Now let $\tilde Q$ be a $N$-by-$N$ matrix, with $\vec e_1,\vec u_1,\cdots,\vec u_{N-1}$ its column vectors, then   
\begin{equation}
\tilde Q^T d\tilde F_{p} \tilde Q = diag(-\tilde f'_{12}(d_{12}) - \tilde f'_{13}(d_{13}),\tilde \lambda_1,\cdots,\tilde \lambda_{N-1})
\end{equation}
The inertia of $d\tilde F_{p}$ thus coincides with the inertia of the diagonal matrix above.  This then proves the lemma.
\end{proof}
\vspace{5pt}

Theorem \ref{ASThm1} is then established by combining lemma \ref{ASlem1} and lemma \ref{ASlem2}.  

%%%%%%%%%%%%%%%%%%%%%%%%%%%%%%%%%%%%%%%%%%%%%%%%%%%%%%%%%
%%%%%%%%%%%%%%%%%%%%%%%%%%%%%%%%%%%%%%%%%%%%%%%%%%%%%%%%%
%%%%%%%%%%%%%%%%%%%%%%%%%%%%%%%%%%%%%%%%%%%%%%%%%%%%%%%%%
\section{Nondegenerate critical orbits of \\ line configurations}
Our goal in this section is to prove the next theorem.  
\vspace{5pt}

\begin{theorem}\label{NCOThm1}
Let $K$ be the collection of critical orbits of line configurations of system \eqref{MODEL}, then generically $K$ is a finite set and each critical orbit in $K$ is nondegenerate.
\end{theorem}
\vspace{5pt}

The proof of theorem \ref{NCOThm1} is divided into two parts. We first show that  $K$ is generically  a finite set, and then we show that generically each critical orbit $\mathcal{O}_p$ in $K$ is nondegenerate.      

\subsection{Proof that $K$ is generically  finite}

We will prove this fact by induction on the number of agents. 
\vspace{3pt}
\\
{\it Base case}. Suppose $N=2$, then there is only one critical orbit  $\mathcal{O}_p$, characterized by the condition that $f_{12}(d_{12}) = 0$. 
\vspace{3pt}
\\
{\it Inductive step}. Suppose the lemma holds for $N = n$ with $n\ge 2$, we then prove for $N = n+1$. 
\vspace{3pt}

%Choose a Henneberg construction of $\mathbb{G}$, and again we assume that vertex $1$ is last vertex joining to the graph, via edges $(1,2)$ and $(1,3)$ to vertices $2$ and $3$. %We then consider all three reduced systems. 
%Let  $p^*$ be the sub-configuration of $p$, formed by agents $\vec x_2,\cdots,\vec x_N$, then $p^*$ is an equilibrium of the reduced system associated with $p$. 
%\vspace{3pt}

We recall that $\Omega^*$ is the collection of ensembles of interaction laws in a reduced system. Let $\Omega^*_1\subset \Omega^*$ be defined by collecting ensembles of interaction laws by which there are only finitely many critical orbits of line configurations.  Then by induction, the set $\Omega^*_1$ contains an open dense subset of $\Omega^*$, and for simplicity, we may assume that $\Omega^*_1$ itself is open and dense in $\Omega^*$. 
\vspace{3pt}

Let $\Omega_1$ be a subset of $\Omega$ defined in the same way as $\Omega^*_1$ in $\Omega^*$, yet with respect to the original system. Let $\rho_1$, $\rho_2$ and $\rho_3$ be maps from $\Omega$ to $\Omega^*$ defined at the end of section \ref{VIL}. We then consider a set
\begin{equation}
\Omega'_1:= \bigcap^3_{i=1}\rho^{-1}_i \Omega^*_1
\end{equation}
It is clear that $\Omega_1$ contains $\Omega'_1$ as a subset. So it suffices to show that $\Omega'_1$ is open  dense in $\Omega$.
\vspace{3pt}

First we show that $\Omega'_1$ is open in $\Omega$. Each $\rho^{-1}_i\Omega^*_1$ is open in $\Omega$ because $\rho_i$ is continuous, and so is their intersection. We now show that  $\Omega'_1$ is dense in $\Omega$.  Suppose not, there is an element $\omega$ in $\Omega$ and an open neighborhood $U$ of $\omega$ in $\Omega$ such that $U\cap \Omega'_1 = \varnothing$. Then there must be some $\rho_i$ with $\rho_i(U) \cap \Omega^*_1 = \varnothing$. On the other hand, the set $\rho_i(U)$ is open in $\Omega^*$ because $\rho_i$ is an open map, and $\Omega^*_1$ is dense in $\Omega^*$, so $\rho_i(U)$ must intersect $\Omega^*_1$ which is a contradiction.

\subsection{Proof that  each $\mathcal{O}_p$ in $K$ is generically nondegenerate}

It suffices to show that generically $n_0(H_p) = 3$ for all $\mathcal{O}_p$ in $K$.  We again prove this fact by induction on the number of agents. 
\vspace{3pt}
\\
{\it Base case}. Suppose $N=2$, then there is only one critical orbit $\mathcal{O}_p$. Suppose $p$ is on the $a$-axis, then
we have 
\begin{equation}
F_p = 
\begin{pmatrix}
0 & 0\\
0 & 0
\end{pmatrix}
\hspace{10pt} \text{and} \hspace{10pt}
d\tilde F_{p} = \tilde f'_{12}(d_{12}) 
\begin{pmatrix}
-1 & 1\\
1 & -1
\end{pmatrix}
\end{equation}  
and hence 
$
n_0(H_p) = n_0(F_p) + n_0(d\tilde F_p) = 2 + 1 = 3 
$. 
\vspace{5pt}
\\
{\it Inductive step}. Suppose the lemma holds for $N = n$ with $n\ge 2$,  we then prove for $N = n+1$. 
\vspace{3pt}

Let $\Omega^*_2$ be a subset of $\Omega^*_1$ defined by  collecting ensembles of interaction laws  by which, we have $n_0(H_{p^*}) = 3$ for each critical orbit $\mathcal{O}_{p^*}$ of line configurations. By induction, the subset $\Omega^*_2$ contains an open dense subset of $\Omega^*$, and we again assume that $\Omega^*_2$ itself is open and dense in $\Omega^*$. Similarly, we let $\Omega_2$ be a subset of $\Omega_1$ defined in the same way as $\Omega^*_2$ in $\Omega^*_1$, yet with respect to the original system. Let
\begin{equation}
\Omega'_2 := \bigcap^3_{i=1} \rho^{-1}_i \Omega^*_2 
\end{equation} 
then by the same reason as $\Omega'_1$ in $\Omega$, the set $\Omega'_2$ is also open dense in $\Omega$. Let 
\begin{equation}
\Omega''_2 := \Omega_2 \cap \Omega'_2 
\end{equation}
It then suffices to show that $\Omega''_2$ is open and dense in $\Omega'_2$.  
\vspace{3pt}
\\
{\it i) Openness}.  We first show that $\Omega''_2$ is open in $\Omega$. Let $\omega$ be an element in $\Omega''_2$, and let $\mathcal{O}_{p_1},\cdots, \mathcal{O}_{p_k}$ be the critical orbits in $K$ associated with $\omega$. Then for each $i=1,\cdots, k$, there is an open neighborhood $U_i$ of $\omega$ in $\Omega$, and an open neighborhood $V_i$ of $\mathcal{O}_{p_i}$ in $P$ such that if $\omega'$ is in $U_i$, then there is a unique nondegenerate critical orbit $\mathcal{O}_{p'_i}$ in $V_i$. Moreover, the critical orbit $\mathcal{O}_{p'_i}$ is an orbit of line configurations. So if we let $U:=\cap^k_{i=1}U_i$, then $U$ is an open subset of $\Omega$ contained in $\Omega''_2$.  
\vspace{3pt}
\\
{\it ii) Density}. We now show that $\Omega''_2$ is dense in $\Omega'_2$. Let $\omega$ be an element in $\Omega'_2$, we show that any open neighborhood $U$ of $\omega$ in $\Omega$ intersects $\Omega''_2$.  Let $\mathcal{O}_{p_1},\cdots,\mathcal{O}_{p_k}$ be the critical orbits in $K$ associated with $\omega$. Let $p^*_i$ be the sub-configuration of $p_i$ formed by agents $\vec x_2,\cdots,\vec x_{n+1}$, then each $\mathcal{O}_{p^*_i}$ is a nondegenerate critical orbit of the reduced system associated with $p_i$.  By the same arguments as we used to prove the openness of $\Omega''_2$ in $\Omega$, we conclude that  there is an open neighborhood $U$ of $\omega$ in $\Omega$,  and an open neighborhoods $V_i$ of each $\mathcal{O}_{p_i}$ in $P$ such that if $\omega'$ is in $U$, then there is a unique critical orbit of line configuration $\mathcal{O}_{p'_i}$ in $V_i$. Moreover, each $\mathcal{O}_{p'^*_i}$ is a nondegenerate critical orbit of the reduced system associated with $p'_i$. In the rest of the proof, we will fix this open neighborhood $U$, and we assume that any perturbation of  $\omega$ is made within $U$. 
\vspace{3pt}

Let $W_i$ be a subset of $U$ collecting any $\omega'$ by which $n_0(H_{p'_i}) = 3$.  It then suffices to show that any open neighborhood $U'$ of $\omega$ in $U$ intersects $W_i$ for each $i=1,\cdots,k$.  By theorem \ref{ASThm1}, we know 
\begin{equation}
3\le n_0(H_{p_i})\le 5
\end{equation} 
and we may assume the worst case where $\mathcal{O}_{p_i}$ is degenerate with $n_0(H_{p_i}) = 5$, i.e,  
\begin{equation}
\left\{
\begin{array}{l}
f_{12}(d_{12}) + f_{13}(d_{13}) = 0 \vspace{3pt}\\
\tilde f'_{12}(d_{12}) + \tilde f'_{13}(d_{13}) =  0
\end{array}
\right.
\end{equation} 
We will now perturb $\omega$ to an $\omega'$  in $W_i\cap U'$. For simplicity,  we only  focus on the case where  agent $\vec x_1$ is in between $\vec x_2$ and $\vec x_3$ in $p_i$, but the analysis will be the same in the other two cases. Choose a triplet $(\epsilon_{12},\epsilon_{13},\epsilon_{23})$ in $\Pi^3_{i=1}C^1(\mathbb{R}_+, \mathbb{R})$ such that they satisfy the condition
\begin{equation}
\left\{
\begin{array}{l}
\tilde \epsilon_{12}(d_{12}) = \tilde \epsilon_{13}(d_{13}) = -\tilde \epsilon_{23}(d_{23}) >0  \vspace{3pt}\\
\tilde \epsilon'_{12}(d_{12}) + \tilde \epsilon'_{13}(d_{13}) >0 
\end{array}
\right.
\end{equation}   
and we assume that $(\epsilon_{12},\epsilon_{13},\epsilon_{23})$ are small enough such that 
\begin{equation}
\omega':= \omega + ( \epsilon_{12}, \epsilon_{13},  \epsilon_{23}, 0,\cdots,0)
\end{equation}
is contained in $U'$. Then by construction, the perturbed critical orbit $\mathcal{O}_{p'_i}$ coincides with the original one $\mathcal{O}_{p_i}$. But by this perturbation, we have
\begin{equation}
\begin{array}{ll}
& \big (f_{12}(d_{12}) +\epsilon_{12}(d_{12})\big)+\big( f_{13}(d_{13}) + \epsilon_{13}(d_{13})\big) \vspace{3pt}\\
= & \tilde \epsilon_{12}(d_{12})/d_{12} + \tilde \epsilon_{13}(d_{13})/d_{13} >0
\end{array}
\end{equation}
and also 
\begin{equation}
\begin{array}{ll}
& \big (\tilde f'_{12}(d_{12}) + \tilde \epsilon'_{12}(d_{12})\big)+\big( \tilde f'_{13}(d_{13}) + \tilde \epsilon'_{13}(d_{13})\big) \vspace{3pt}\\
 = &  \tilde \epsilon'_{12}(d_{12}) + \tilde \epsilon'_{13}(d_{13}) > 0
 \end{array}
\end{equation}
So then $n_0(H_{p'_i}) = 3$ after perturbation, and hence $\omega'$ is contained in $W_i$. This then completes the proof.

%%%%%%%%%%%%%%%%%%%%%%%%%%%%%%%%%%%%%%%%%%%%%%%%%%%
%%%%%%%%%%%%%%%%%%%%%%%%%%%%%%%%%%%%%%%%%%%%%%%%%%%
%%%%%%%%%%%%%%%%%%%%%%%%%%%%%%%%%%%%%%%%%%%%%%%%%%%
%%%%%%%%%%%%%%%%%%%%%%%%%%????%%%%%%%%%%%%%%%%%%%%%

\section{Proof of the main theorem}
Let $\mathbb{G} = (V,E)$ be a TLG of $N$ vertices, and let $S_{\mathbb{G}}$ be the collection of all triangulated Laman subgraphs of $\mathbb{G}$. Let $\mathbb{G}'= (V',E')$ be an element in $S_{\mathbb{G}}$, and we consider a sub-system induced by $\mathbb{G}'$, i.e, the sub-system formed by agents $\vec x_i$ with $i\in V'$, and $\{f_{jk}| (j,k) \in E'\}$ is the ensemble of interaction laws. Let 
\begin{equation}
\Omega_{\mathbb{G}'} := \mathcal{F}^{|E'|}  
\end{equation} 
be the collection of ensembles of interaction laws associated with the sub-system. Let $\hat \Omega_{\mathbb{G}'}\subset \Omega_{\mathbb{G}'}$ be the subset collecting any ensemble of interaction laws by which the associated potential function of the sub-system is an equivariant Morse function. By theorem \ref{NCOThm1}, the set $\hat\Omega_{\mathbb{G}'}$ is open dense in $\Omega_{\mathbb{G}'}$. Now for convenience, we let 
\begin{equation}
\Omega_{-\mathbb{G}'} := \mathcal{F}^{|E|-|E'|} 
\end{equation}
and we define a subset of $\Omega$ by
\begin{equation}
\hat\Omega:= \bigcap_{\mathbb{G}'\in S_{\mathbb{G}}}\big ( \hat\Omega_{\mathbb{G}'} \times \Omega_{-\mathbb{G}'} \big)
\end{equation}
Since $\hat\Omega$ is a finite  intersection of open dense subsets of $\Omega$, and hence $\hat\Omega$ itself is open dense in $\Omega$. 
\vspace{3pt}

It now suffices to show that for each element $\omega$ in $\hat\Omega$, the associated potential function is an equivariant Morse function. Let $\mathcal{O}_p$ be a critical orbit,  and let $\{p_i\}^m_{i=1}$ and $\{\mathbb{G}_i\}^m_{i=1}$ be sub-configurations of $p$ and sub-graphs of $\mathbb{G}$ respectively,  associated with the canonical partition. Then each $\mathcal{O}_{p_i}$ is a critical orbit of line configurations of the sub-system induced by $\mathbb{G}_i$. As each $\mathbb{G}_i$ is a TLG, so by construction of $\hat\Omega$, we have  $n_0(H_{p_i}) = 3$ for all $i=1,\cdots,m$. We then apply the Morse-Bott index formula to conclude that  $n_0(H_p) = 3$.  In other words, we have just showed that  each critical orbit  $\mathcal{O}_p$ is nondegenerate. As there are only finitely many critical orbits, so the associated potential function is an equivariant Morse function.

\section*{Appendix}
We here will prove that each map $\rho_i$, with $i=1,2,3$,  defined at the end of section \ref{VIL} is open, surjective and continuous. For simplicity, we will only focus on the proof about $\rho_1$,  the same arguments can be applied to the other two cases. 
\vspace{10pt}
\\
{\it 1. Proof that $\rho_1$ is surjective}
\vspace{5pt}

Let $\omega^*$ be an element in $\Omega^*$, and we now show that there is  an element $\omega$ in $\Omega$ with $\rho(\omega) = \omega^*$. By construction, it suffices to show that for any function $ f^*_{23}$ in $\mathcal{F}$, there is a triplet $(f_{12}, f_{13}, f_{23})$ in $\mathcal{F}^3$ such that 
\begin{equation}\label{APXeq1}
f^*_{23} = f_{23} + g_{23} 
\end{equation}
with $g_{23}$ the virtual interaction induced by $f_{12}$ and $f_{13}$. Let
\begin{equation}
\left\{
\begin{array}{l}
f_{23}(d) :=  \frac{1}{2} f^*_{23}(d)\vspace{3pt} \\ 
f_{12}(d) = f_{13}(d)  := \frac{1}{2}f^*_{23}(2d)
\end{array}
\right.
\end{equation}
for all $d>0$. We then verify that all $f_{12}$, $f_{13}$ and $f_{23}$ are in $\mathcal{F}$, and $g_{23} = f^*_{23}/2$ is the virtual interaction induced by $f_{12}$ and $f_{13}$. Thus, the expression \eqref{APXeq1} is satisfied by our choice of $f_{12}$, $f_{13}$ and $f_{23}$. 
\vspace{10pt}
\\
{\it 2. Proof that $\rho_1$ is open}
\vspace{5pt}

Let $U$ be an open set in $\Omega$, let $U^*:= \rho_1(U)$ and we show that $U^*$ is an open set in $\Omega^*$. Let $\omega^*$ be  in $U^*$, and let $\omega = (\cdots, f_{ij},\cdots)$ be in $U$ with $\rho_1(\omega) =\omega^*$. Choose open neighborhoods $V_{ij}$ of $f_{ij}$ in $\mathcal{F}$ such that $\Pi_{(i,j)\in E}V_{ij}$ is in $U$.  Let 
\begin{equation}
E^*:= E - \{(1,2), (1,3)\}
\end{equation}
and let  $U_1$ be a subset of $U$ defined by
\begin{equation}
U_1 := f_{12}\times f_{13} \times \Pi_{(i,j)\in E^*} V_{23} 
\end{equation}
Let $U^*_1 := \rho_1(U_1)$, then we have
\begin{equation}
U^*_1:=  \Pi_{(i,j)\in E^*} V_{ij} -  \Pi_{(i,j)\in E^*} f_{ij} + \omega^* 
\end{equation}
So $U^*_1$  is an open neighborhood of $\omega^*$ contained in $U^*$. 
\vspace{10pt}
\\
{\it 3. Proof that $\rho_1$ is continuous}
\vspace{5pt}

 Let $\eta: \mathcal{F}^2\to \mathcal{F}$ be the map defined by sending a pair $(f_{12},f_{13})$ to the virtual interaction $g_{23}$, it then suffices to show that  $\eta$ is continuous. The proof will be divided into two parts, we will first fix a distance $d>0$, and prove that the map $\eta_d: \mathcal{F}^2\to \mathbb{R}^2$ defined by
 \begin{equation}
 \eta_d: (f_{12},f_{13})\mapsto (g_{23}(d),g'_{23}(d))
 \end{equation}
 is continuous. The continuity of $\eta$ will then be established afterwards.
\vspace{5pt}
\\
{\it i) Continuity of $\eta_d$}
\vspace{3pt}

Let $(f_{12},f_{13})$ be a pair contained in $\mathcal{F}^2$, and let $(d_{12},d_{13})$ be the unique pair of positive numbers with
\begin{equation}\label{d1213}
\left\{
\begin{array}{l}
d_{12} + d_{13} = d \vspace{3pt}\\
\tilde f_{12}(d_{12}) = \tilde f_{13}(d_{13}) 
\end{array}
\right.
\end{equation}
It is clear that for fixed $d$, the pair $(d_{12},d_{13})$ is a function of $(f_{12},f_{13})$ only.  Notice that 
\begin{equation}
\left\{
\begin{array}{l}
\tilde g_{23}(d) = \tilde f_{12}(d_{12})\vspace{3pt}\\
\tilde g'_{23}(d) = \frac{\tilde f'_{12}(d_{12})\tilde f'_{13}(d_{13})}{\tilde f'_{12}(d_{12}) + \tilde f'_{13}(d_{13})} 
\end{array}\right.
\end{equation}
So it suffices to show that the pair $(d_{12},d_{13})$ is continuous in $(f_{12},f_{13})$. We shall use the inverse function theorem to establish this fact. 
\vspace{3pt}

As  continuity is a local problem, so it suffices to consider variations of $f_{12}$ and $f_{13}$ over some neighborhoods of $d_{12}$ and $d_{13}$. Let $I$ be any closed neighborhood of $0$ in $\mathbb{R}$, and let $I_{12} $ and $I_{13}$ be closed intervals defined by
\begin{equation}\label{I1213}
\left\{
\begin{array}{l}
I_{12}:= d_{12} + I \vspace{3pt}\\
I_{13}:= d_{13} -  I
\end{array}
\right. 
\end{equation} 
%So then $I_{12}$ and $I_{13}$ are closed neighborhoods of $d_{12}$ and $d_{13}$ respectively. %As the continuity of $(d_{12},d_{13})$ is a local problem, so we only need to consider variations of $f_{12}$ and $f_{13}$ when restricted to ${I_{12}}$ and ${I_{13}}$ respectively. 
So then $I_{12}$ and $I_{13}$ are closed neighborhoods of $d_{12}$ and $d_{13}$ respectively. We now consider a $C^1$-map 
\begin{equation}
\xi:  I\times C^1(I_{12},\mathbb{R})\times C^1(I_{13},\mathbb{R})\to \mathbb{R}
\end{equation} 
defined  by 
\begin{equation}
\xi(x,h_{12}, h_{13})  :=  \tilde h_{12}(d_{12} + x) - \tilde h_{13}(d_{13} - x)  
\end{equation}
Let $f_{12}|_{I_{12}}$ and $f_{13}|_{I_{13}}$ be restrictions of $f_{12}$ and $f_{13}$ to $I_{12}$ and $I_{13}$ respectively, then we have 
\begin{equation}
\left\{
\begin{array}{l}
\xi(f_{12}|_{I_{12}},f_{13}|_{I_{13}},0) = 0 \vspace{3pt}\\
\frac{\partial\xi} {\partial x}\Big |_{(f_{12}|_{I_{12}},f_{13}|_{I_{13}},0)} = \tilde f'_{12}(d_{12}) + \tilde f'_{13}(d_{13}) > 0
\end{array}
\right.
\end{equation}
So by the inverse function theorem for the Banach spaces, we know that there exist
\vspace{2pt}
\begin{itemize}
\item[1).] an open neighborhood $V_{12}$ of $f_{12}|_{I_{12}}$ in $C^1(I_{12},\mathbb{R})$;
\vspace{2pt}
\item[2).] an open neighborhood $V_{13}$ of $f_{13}|_{I_{13}}$ in $C^1(I_{13},\mathbb{R})$;
\vspace{2pt}
\item[3).] an open interval $(-t, t)$, with $t>0$, in $\mathbb{R}$;
\vspace{2pt}
\item[4).] a unique $C^1$-map $\sigma: V_{12} \times V_{13} \to (-t,t) $
\end{itemize}
\vspace{2pt}
such that 
\begin{equation}
\xi(h_{12},h_{13},\sigma(h_{12},h_{13})) = 0
\end{equation} 
for all $(h_{12},h_{13})$ in $V_{12}\times V_{13}$. This then shows the continuity of $(d_{12},d_{13})$ at $(f_{12},f_{13})$. 
\vspace{5pt}
\\
{\it ii) Continuity of $\eta$}
\vspace{3pt}

Let $(f_{12},f_{13})$ be any pair in $\mathcal{F}^2$, we will prove the continuity of $\eta$ at $(f_{12},f_{13})$. 
As the distance $d$ is no longer fixed, so to avoid confusion, we will write the argument $d$ explicitly out, e.g, we will write $d_{12}(d)$ and $d_{13}(d)$ instead of $d_{12}$ and $d_{13}$.
\vspace{3pt} 

We start by defining a family of closed intervals $I(d)$, parametrized by $d$ over $\mathbb{R}_+$. Let
\begin{equation}
I(d):=[-\alpha(d), \beta(d)]
\end{equation} 
with both $\alpha$ and $\beta$ functions in $C^0(\mathbb{R}_+)$, and  we let
\begin{equation}
\left\{
\begin{array}{l}
I_{12}(d):=  d_{12}(d) + I(d)\vspace{3pt}\\
I_{13}(d):=  d_{13}(d) - I(d)   
\end{array}
\right.
\end{equation}  
be closed neighborhoods of $d_{12}(d)$ and $d_{13}(d)$ respectively. Now for each $d>0$, we define two subsets of $\mathbb{R}_+$ by
\begin{equation}
\left\{
\begin{array}{l}
J_{12}(d) := \{d'\in \mathbb{R}_+| d\in I_{12}(d')\}\vspace{3pt}\\
J_{13}(d) := \{d'\in\mathbb{R}_+| d\in I_{13}(d')\}
\end{array}
\right.
\end{equation}
We then assume that $\alpha$ and $\beta$ are chosen such that 
\vspace{2pt}
\begin{itemize}
\item[1.] both $I_{12}(d)$ and $I_{13}(d)$ are in $\mathbb{R}_+$;
\vspace{3pt}
\item[2.] both $J_{12}(d)$ and $J_{13}(d)$ are compact
\end{itemize}
\vspace{2pt}
 for all $d>0$.
\vspace{3pt}

We will now show that for any $\delta$ in $C^0(\mathbb{R}_+)$, there exists an $\epsilon$ in $C^0(\mathbb{R}_+)$ such that 
\begin{equation}
B_{\epsilon}(f_{12})\times B_{\epsilon}(f_{13}) \subset \eta^{-1}(B_{\delta}(g_{23})) 
\end{equation}  
First by the proof of continuity of $\eta_d$, we know that for a fixed $d>0$, there exists a number $r_d>0$ such that if 
\begin{equation}
\max\big \{|\epsilon(x)| + |\epsilon'(x)| \big | x\in I_{12}(d)\cup I_{13}(d)\big\} < r_d
\end{equation}
then for any $(h_{12},h_{13})$ in $B_{\epsilon}(f_{12})\times B_{\epsilon}(f_{13})$, we have
\begin{equation}
\big | \eta_d(h_{12}, h_{13}) - \eta_d(f_{12},f_{13})\big |_1 < \delta(d) 
\end{equation} 
where $|\cdot|_1$ denotes the $1$-norm of a vector in $\mathbb{R}^2$. Let 
\begin{equation}
\hat r_{d}:= \min \big \{r_{d'}\big| d'\in J_{12}(d)\cup J_{13}(d) \big\} 
\end{equation}
As $J_{12}(d)$ and $J_{13}(d)$ are compact, so then $\hat r_{d}>0$. Now we choose an $\epsilon$ in $C^0(\mathbb{R}_+)$ such that 
$
\epsilon(d)  < \hat r_d 
$ 
for all $d>0$. This then  implies that 
\begin{equation}
\eta(h_{12},h_{13}) \in B_{\delta}(g_{23})
\end{equation}
for all $(h_{12},h_{13})$ in $B_{\epsilon}(f_{12})\times B_{\epsilon}(f_{13})$.

 \bibliographystyle{unsrt}
\bibliography{FC}

\begin{thebibliography}{1}

\bibitem{chen2014reciprocal}
X.~Chen, M.-A. Belabbas, and T.~Ba\c{s}ar.
\newblock Formation control with triangulated {L}aman graphs.
\newblock {\em arXiv preprint arXiv:1412.6958}, 2014.

\bibitem{UH2013E}
U.~Helmke and B.D.O. Anderson.
\newblock Equivariant {M}orse theory and formation control.
\newblock In {\em Communication, Control, and Computing (Allerton), 2013 51st
  Annual Allerton Conference on}, pages 1576--1583. IEEE, 2013.

\bibitem{AB2012CDC}
M.-A. Belabbas, S.~Mou, A.S. Morse, and B.D.O. Anderson.
\newblock Robustness issues with undirected formations.
\newblock In {\em Conference on Decision and Control (CDC), 2012}, pages
  1445--1450. IEEE, 2012.

\bibitem{sun2014CDC}
Z.~Sun, S.~Mou, B.D.O. Anderson, and A.S. Morse.
\newblock Formation movements in minimally rigid formation control with
  mismatched mutual distances.
\newblock In {\em Conference on Decision and Control (CDC), 2014}. IEEE, 2014.

\bibitem{USZB}
U.~Helmke, S.~Mou, Z.~Sun, and B.D.O Anderson.
\newblock Geometrical methods for mismatched formation control.
\newblock In {\em The 53rd Conference on Decision and Control(CDC), 2014}.
  IEEE, 2014.

\bibitem{chenRMAS}
X.~Chen.
\newblock Swarm aggregation in reciprocal multi-agent systems with fading
  interaction laws.
\newblock {\em arXiv preprint arXiv:1412.6952}, 2014.

\bibitem{laman1970}
G.~Laman.
\newblock On graphs and rigidity of plane skeletal structures.
\newblock {\em Journal of Engineering Mathematics}, 4(4):331--340, 1970.

\bibitem{sylvester1852xix}
J.J. Sylvester.
\newblock Xix. a demonstration of the theorem that every homogeneous quadratic
  polynomial is reducible by real orthogonal substitutions to the form of a sum
  of positive and negative squares.
\newblock {\em The London, Edinburgh, and Dublin Philosophical Magazine and
  Journal of Science}, 4(23):138--142, 1852.

\end{thebibliography}

\end{document}